\documentclass[journal]{IEEEtran}
\usepackage{cite}
\ifCLASSINFOpdf
\else
\fi
\usepackage{microtype}
\usepackage{array}
\usepackage{mdwmath}
\usepackage{mdwtab}
\usepackage{graphicx}
\usepackage{euscript}
\usepackage{amsfonts}
\usepackage{bigstrut}
\usepackage{tabularx}
\usepackage{cite}
\usepackage{amsmath}
\usepackage{algorithm}
\usepackage[noend]{algpseudocode}
\makeatletter
\def\BState{\State\hskip-\ALG@thistlm}
\makeatother
\usepackage{enumitem}
\usepackage{upgreek}
\usepackage{dsfont}
\usepackage{amsthm}
\usepackage{amssymb}
\usepackage{textcomp}
\usepackage{gensymb}
\usepackage{xcolor}
\usepackage{mathtools}

\usepackage{hyperref}
\usepackage{epstopdf} 
\usepackage{bbm}
\DeclareMathOperator*{\argmax}{argmax} 
\usepackage{microtype}
\usepackage{booktabs}

\usepackage{multirow}

\title{
HybNet: A Hybrid Deep Learning - Matched Filter Approach for IoT Signal Detection
}

\author{Kosta Dakic$^\star$,~\IEEEmembership{Graduate~Student Member,~IEEE}, Bassel~Al~Homssi,~\IEEEmembership{Member,~IEEE},\\ Margaret~Lech,~\IEEEmembership{Member,~IEEE}, and Akram Al-Hourani$^\star$,~\IEEEmembership{Senior~Member,~IEEE},
\thanks{$^\star$Corresponding authors: K.~Dakic and A. Al-Hourani, with the School of Engineering, RMIT University, Melbourne, VIC 3000. E-mail: kosta.dakic@ieee.org, akram.hourani@rmit.edu.au.}
 } 

\begin{document}

\maketitle

\begin{abstract}
Random access schemes are widely used in IoT wireless access networks enabling reduced complexity and overcoming power consumption constraints. Nevertheless, random access results in high packet losses which are caused by overlapping transmissions. Signal detection methods for digital modulation techniques are typically based on the well-established \textit{matched filter}, which is proven as the optimal filter under additive white Gaussian noise for minimizing error probability. However, with the \emph{colored} interference arising from the overlapping IoT transmissions, deep learning approaches are being considered a suitable alternative. In this paper, we present a hybrid framework, dubbed as \emph{HybNet}, that alternates between deep learning and match filter pathways based on the perceived interference level. This helps the detector to work in a broader range of conditions, optimally leveraging both the matched filter and deep learning advantages. We compare the performance of several possible data modalities and detection architectures concerning the interference-to-noise ratio, demonstrating that leveraging domain knowledge by pre-processing the input data in conjunction with the proposed HybNet surpasses the complex conjugate matched filter performance under interference-limited scenarios.
\end{abstract}

\section{Introduction}~\label{Section_I}
Internet-of-Things (IoT) wireless technologies play a significant role by facilitating the connectivity between sensor devices and Internet cloud services. According to a report from Ericsson, over 5 billion cellular IoT connections are anticipated by 2025~\cite{mobility_report}. For wireless IoT devices, this increase would further congest the scarce radio spectrum and increase the level of interference. This is already evident in the license-free \textit{industrial, scientific, and medical} (ISM) band~\cite{8403749} since this band is available to the public and does not require paid licensing. 

In centrally-controlled networks, the base station, e.g., gNB in 5G, orchestrates the access to the spectral resources by employing a multiple-access technique such as time-division multiple access (TDMA), frequency-division multiple access (FDMA), code-division multiple access (CDMA), and orthogonal frequency-division multiple access (OFDMA). These access techniques are required to alleviate the co-channel interference. Nevertheless, such techniques mandate significant overhead, adding complexity to the protocol stack. Thus, typical license-free IoT networks employ random access techniques to simplify the protocol stack, thereby lowering energy consumption as well as device cost. Furthermore, a complex access technique would not be efficient in the first place given that the interference is originating from other co-existing systems in the same band.

Efficient signal detection is crucial for Low power wide area networks (LPWAN) technologies, which are receiving increasing attention due to the increasing use of IoT applications. LPWANs can achieve long-distance communication while maintaining low-power consumption at the cost of a reduced bit rate. In a license-free spectrum, LPWAN network performance is typically an interference-limited system in urban environments, while it becomes noise-limited in rural and remote locations. A prominent LPWAN technology is LoRaWAN, adopted by the LoRa Alliance~\cite{lorawan}. LoRaWAN typically uses an ALOHA-based protocol, allowing multiple IoT devices to transmit without coordination or prior signaling. However, the transmitted signals are prone to packet collisions which significantly reduces the throughput and in turn, the network performance~\cite{RN590,9239466,9395074,RN308,8581011}. LoRaWAN utilizes the LoRa modulation scheme, which is based on chirp spread spectrum (CSS) modulation spreading the signal energy over a wider bandwidth to combat narrowband interference~\cite{RN308}. Furthermore, the time-spreading of the transmission can be controlled using a parameter called \emph{spreading factor} (SF), which increases the energy at the receiver without the need to transmit at a higher RF power. Aside from potential narrowband interference, LoRa-to-LoRa-induced interference significantly degrades the performance, especially when the interfering LoRa signal uses the same SF. 

To achieve signal (frame) detection, receivers typically employ matched filter architecture, where the signal is compared, or \emph{matched}, to a known template(s). This method is proven to be optimal under additive white Gaussian noise (AWGN) conditions~\cite{simon2005digital}. However, the rapid evolution of neural networks (NN), particularly deep learning (DL) methods, has shown great potential for signal detection~\cite{RN525,RN318,RN558,RN565,AIpaper}. The main appeal of DL signal detection is its strength against non-linearities~\cite{RN30}. Notably, the Convolutional Neural Network (CNN) has been utilized to classify stochastic signals such as in image classification~\cite{GU2018354}, which is not practically possible using deterministic methods such as the matched filter.
 
To our knowledge, there has not been any previous research work focusing on mitigating the issue of LoRa modulation with the same technology interference by utilizing DL-based detection. This paper presents a new framework, \emph{HybNet}, that alternates between a (i) matched filter detector and (ii) a proposed DL detector, where the switching is automatically decided based on the interference level. Consequently, Our framework harnesses the benefits of both (i) the matched filter's ideal performance under noise-limited scenarios from one side, and (ii) the improved performance of the proposed DL detector in LoRa-to-LoRa interference on the other side. Additionally, we explore three different input data modalities for the DL-based detection and compare their performance; (i) time-domain (I/Q samples), (ii) time-frequency domain (spectrogram), and (iii) frequency domain (spectrum). Results show that the proposed DL-detector outperforms traditional non-coherent detection in non-Gaussian interference scenarios and without perfect phase synchronization. Also, the frequency domain DL detector shows a greater performance with a lower network complexity compared to the time-domain and time-frequency domain DL detectors due to the smaller input frame size compared to the I/Q frame and the 2D spectrogram. To evaluate the proposed framework at different interference-to-noise levels, we vary the interference-to-noise ratio (INR) and obtain the corresponding bit-error-rate (BER). Accordingly, the overall BER performance results indicate that the proposed DL-based detectors can significantly improve LoRa detection performance for LoRa symbols under high interference. The contribution of this work is summarized as follows.
\begin{itemize}
    \item It presents a novel LoRa symbol detector, \textit{HybNet}, which combines both matched filter detection and DL-enabled detection via an automatic switching mechanism. Through this combination, HybNet can capitalize on matched filter detection capabilities in noisy mediums while benefiting from DL detection under interference.
    \item It demonstrates the compatibility of the proposed neural network with conventionally deployed energy detectors to switch between DL and matched filter detection based on the measured interference level. This provides a trade-off in terms of complexity and detection performance.
    \item It assesses the performance of various data modalities and their corresponding neural network architectures. The data modalities assessed are time, frequency, and time-frequency.
    \item It proposes an experimental framework for systematically evaluating symbol detection efficiency using various input data modalities, AWGN impairments, and interference levels. 
\end{itemize}

The rest of this paper is organized as follows. A literature review is presented in Section~\ref{Section_II}. An overview of the systems model which utilizes LoRa modulation along with LoRa emulation and dataset creation is covered in Section~\ref{Section_III}. Section~\ref{Section_IV} discusses the detection architectures utilized in this paper. Section~\ref{Section_V} discusses the results obtained from the experiments. Finally, Section~\ref{Section_VI} concludes the paper.

\section{Background and Related work}~\label{Section_II}
This section summarizes some of the  previous works. This includes those that addressed various ways to mitigate \emph{colored} interference, which occurs due to the overlapping of the transmitted signals in time-frequency. Moreover, it includes works on various machine learning (ML) and DL techniques that have been applied to wireless communication for signal detection. 

\subsection{Interference and LoRa}
Interference is a significant challenge that hinders adequate packet reception, particularly in shared radio frequency bands. In these bands, different users and systems are allowed to access the spectrum without prior resource coordination to simplify their operation, ultimately decreasing spectrum efficiency. Thus, extensive work has been conducted on how to efficiently and dynamically utilize the available spectrum relying on techniques such as cognitive radio (CR). CR relies on being able to sense the spectrum occupancy and devise a spectrum access plan to mitigate co-channel interference~\cite{RN232}. However, the efficiency of CR is subject to accuracy in estimating  the occupancy in the spectrum. As such, random access networks severely reduce such capacity. Also, CR requires additional computational resources (and hence power) for spectrum sensing and prediction, which is not ideal for battery-operated devices~\cite{RN564}. Hence, CR has not been adopted in IoT networks due to these limitations. 

Another technique to increase the capacity of a system with interference from another signal is to utilize successive interference cancellation (SIC)~\cite{RN597}. In SIC, signals coming from different users are successively decoded, with each decoded signal being subtracted from the total received signal. SIC is also utilized in non-orthogonal multiple access (NOMA) methods for next-generation wireless communications. However, the channel gain and time synchronization information of all users should be known by the base station (BS) to achieve an accurate decomposition of the superimposed signals, which entails additional overhead and is often impossible to achieve in shared frequency bands~\cite{RN598}.

As one of the widely adopted technologies, LoRa modulation is claimed to be robust to interference, whereby LoRa modulation provides a certain level of orthogonality in different SF transmissions, owing to the reduced cross-energy among the mismatched LoRa transmissions with different SFs~\cite{RN66}. However, collisions of inter-SF signals would in practice still cause packet losses, shown in~\cite{RN66,RN308,RN332,unknown}. This is further aggravated in cases where the colliding signals having identical SFs, thus requiring a higher signal to interference plus noise ratio (SINR) to be successfully detected by the receiver~\cite{RN308}. The work in~\cite{RN343} proposes a methodology to allocate different SFs to users to mitigate this effect. However, while the proposed method improves the throughput, interference still increases the error probability due to the non-perfect orthogonality. Recent research has also shown how coherent detection methods enhance the performance in same-SF interference scenarios, demonstrated by theoretical approximations and Monte Carlo simulations as shown in~\cite{unknown}. However, the work focuses on same-SF interference and does not explore inter-SF scenarios. Additional work on investigating the theoretical BER performance of traditional LoRa receivers under same-SF interference is explored in~\cite{8581011}, whereas~\cite{8903531} extends the work on same-SF interference to include chirp misalignment.

\subsection{Machine and deep learning in telecommunication}
ML methods, with a focus on DL, have been recently investigated in the field of wireless communications, including works on automatic modulation recognition (AMR)~\cite{RN232,RN9}, occupancy detection in the license-free band~\cite{9205874,RN1}, and optimization of interference management algorithms~\cite{RN555}. Many more applications of ML in communications exist, which are reviewed in~\cite{RN30} for the physical layer of communications, and in~\cite{RN541} for higher layers. Our previous work in~\cite{RN525} explored the use of DL-based signal detection techniques with CNN models for the LoRa modulation scheme under AWGN, synchronization time offset, and carrier frequency offset impairments. The proposed neural network outperformed the conventionally used non-coherent detector. Another research work that utilizes DL sequence detector networks can be found in~\cite{RN318}, constructing a sliding bidirectional recurrent neural network (SBRNN). The SBRNN can take in information from previous symbols, unlike symbol-by-symbol detectors, to combat inter-symbol interference (ISI) for non-LoRa signals. Another advantage of some DL-based detectors is that they do not need channel state information (CSI) which is further demonstrated in~\cite{RN558} proposing a detection method based on the generative adversarial network (GAN). 

Conventional DL-based detectors in past research literature rely on a data-driven approach, where the DL network learns from a large labeled dataset and mathematical models are not necessarily utilized to aid in training. However, DL-based detectors can be further optimized by leveraging domain knowledge for a model-based approach to significantly reduce the network complexity as well as the dataset size required for training~\cite{8715338}. An example of a model-based approach for signal detection in OFDM systems is presented in~\cite{8509622}, where the research work achieves high performance relative to traditional methods and the model-driven DL approach converges quicker requiring fewer parameters compared to the data-driven DL signal detection receiver shown in~\cite{RN324}. Another work presenting model-based DL signal detection is shown in~\cite{RN565}, which uses DL to learn the log-likelihoods and performs Viterbi signal detection. In addition to solid detection performance, the research work also proposes a method to train in real-time and adapt to time-varying channels. Essentially, leveraging some form of domain knowledge in the selection of the architecture by utilizing statistical models, preparing the input data, employing known mathematical processes, or in the selection of the DL architecture can increase the overall performance, reduce the computational complexity, lessen the dependency of a large dataset, and speed up convergence in the training of a DL detector~\cite{RN666}.

To combat non-Gaussian interference, ML can be used to realise efficient frequency band selection techniques for IoT devices~\cite{8755300}. ML for signal detection is also shown to effectively deal with non-Gaussian interference, such as in~\cite{RN324} for the OFDM scheme, demonstrating robustness to non-linear channel impairments, interference, and comparable BER performance to traditional detection methods. However, conventional detection methods outperform the DL methods in a noise-limited scenarios. The HybNet architecture, on the other hand, leverages optimal detection in AWGN conditions with conventional detection, while also performing well in interference-limited conditions by using DL detection techniques. For detecting multiple-input and multiple-output (MIMO) signals with correlated interference, authors in~\cite{RN513} show a DL-based approach using a maximum likelihood detector and a DL network to predict and remove local correlation among the noise in different symbols. Nonetheless, there seems to be no other research work investigating the issue of LoRa with the same technology interference utilizing DL-based detection. 

\section{IoT Signal Model}~\label{Section_III}
This section explains the general structure of LoRa modulation utilized to generate the synthetic data for training the DL neural network. It also details traditional LoRa detection methods utilized as a comparison  benchmark with the proposed DL method.

\subsection{LoRa Modulation}
LoRa CSS modulation is based on linear cyclic chirping within a given bandwidth, denoted as $B$. Each chirp encodes a single symbol with a given duration, denoted as $T_\mathrm{s}$. The bandwidth according to the LoRaWAN protocol can take one of the following values as $B \in \{125,250,500\}$~kHz, whereas the chirp rate of the symbols is controlled by the corresponding SF, where the $\text{SF} \in \{7,8,9,10,11,12\}$ per the LoRaWAN protocol. Based on the SF the number of available symbols is given by $M = 2^\mathrm{SF}$. We follow the same notations as in~\cite{9395074} and~\cite{RN525} to describe LoRa signal modulation where a single symbol (chirp) is represented as follows,
\begin{equation}
s_k(t) = \exp\left(j2\pi \int_{0}^{t} \left[(\beta x + \zeta_k)_{\text{mod}_B} - \frac{B}{2}\right] \mathrm{d}x \right) ,
\end{equation}
where $\zeta_k$ is the shift frequency representing the symbol value as follows,
\begin{equation}
	\zeta_k = m_k\delta_\mathrm{f},
\end{equation}
where $m_k$ is the data symbol value ${m_k\in\{\text{0,1,...,}M-1\}}$, $k$ is the symbol's index, and $\delta_\mathrm{f}$ represents the frequency step between the shifts. The frequency step in LoRa is designed to be equal to the symbol rate itself, i.e.,$\delta_{f}= B/M$. On the other hand, $\beta$ represents the chirp rate given by,
\begin{equation}
	\beta = \frac{f_{\text{high}} - f_{\text{low}}}{T_\mathrm{s}} = \frac{B}{T_\mathrm{s}},
\end{equation}
where $f_{\text{low}}=f_{\text{c}}+\frac{B}{2}$ and $f_{\text{high}}=f_{\text{low}}-\frac{B}{2}$ are respectively the lower and upper frequency bounds of the chirp around the carrier $f_\mathrm{c}$. Accordingly, the sequence of symbols (message) can be expressed as follows,
\begin{equation}
    x(t)=\sum_{k=1}^{K} s_k(t-k T_\mathrm{s}) ,
\end{equation} 
where $K$ is the total number of symbols contained within a message.

\subsection{Conventional LoRa detection}
The detection of LoRa symbols is typically performed in two steps; (i) the symbol is dechirped with the same chirping rate to convert the received symbol into a single tone, accordingly the sequence of symbols will manifest as a multiple frequency shift keying (MFSK) modulation signal, (ii) then in the second step the symbols are detected using a conventional frequency detector.

To dechirp the signal, each symbol in the received LoRa waveform is multiplied by a synchronized and inverted chirp with zero frequency shift. The dechirping pulse train is represented as follows,
\begin{equation}
	s^{\star}(t) = \sum_{k=1}^{K}\exp\left(j\pi B(t - k T_\mathrm{s})-j2\pi\beta (t - k T_\mathrm{s})^2\right) .
\end{equation}
In an ideal channel, each dechirped LoRa symbol can then be expressed as follows,
\begin{equation}
	y(t) = \sum_{k=1}^{K}\exp\left(j2\pi \zeta_k (t-k T_s)\right),
\end{equation}
which is a sequence of single tones, each tone with a frequency offset $\zeta_k$ corresponding to the value of the symbol. Note that this is now a typical MFSK signal. Accordingly, we can use conventional MFSK detection methods to detect the symbols. Two main approaches are typically utilized; (i) non-coherent detection when phase information is not available, and (ii) coherent detection when both the in-phase and quadrature paths are used in the receiver.

\subsubsection{Coherent detection}~\label{Coh_subsection}
Optimal detection of a LoRa symbol in AWGN environments can be achieved using coherent detection. Coherent detection is executed by correlating the dechirped signal with all permissible frequency shifts, then the estimated symbol value is based on the frequency shift that produces the largest real-valued output. MF of an MFSK LoRa symbol is shown as follows,
\begin{equation}
    \hat{m}_{\text{coh}} = \argmax_k\mathfrak{Re}\left[\int_{0}^{\infty} y(t)w_{\mathrm{k}}(t - \tau) d\tau\right] ,
\end{equation}
where $w_{\mathrm{k}}(t) = \exp\left(j2\pi k\delta_{\mathrm{f}} t\right)$ represents a set of harmonics and $k$ is an integer representing the number of discrete frequency steps ${k = \{0,1, ... , M-1\}}$.. Coherent detection can be efficiently implemented with the FFT~\cite{RN516}, as follows,
\begin{equation}
    \hat{m}_{\text{coh}} = \left\lfloor\frac{1}{\delta_\mathrm{f}} \argmax_f\mathfrak{Re}\left[Y(f)\right]\right\rceil.
\end{equation}
Where the coherent detection is proven to be optimal in AWGN environments~\cite{simon2005digital}. However, this requires, the receiver to be phase-locked to the transmitter, by using a phase-locked loop (PLL)~\cite{RN516}. An alternative method to perform signal detection without phase synchronization is to use non-coherent detection.

\subsubsection{Non-coherent detection}~\label{nCoh_subsection}
Non-coherent detection follows the methodology of coherent detection. Whereby each LoRa symbol is dechirped and then correlated with permissible frequency shifts. However, the magnitude of the output is taken, which is shown as follows,
\begin{equation}
    \hat{m}_{\text{ncoh}} = \argmax_k\left|\int_{0}^{\infty} y(t)w_{\mathrm{k}}(t - \tau) d\tau\right| .~\label{nCoh_corr}
\end{equation}
Equivalently, the FFT can be applied as follows,
\begin{equation}
    \hat{m}_{\text{ncoh}} = \left\lfloor\frac{1}{\delta_\mathrm{f}} \argmax_f|Y(f)|\right\rceil,
\end{equation}
where $Y(f) = \text{FFT}\{y(t)\}$ denotes the FFT of the received FSK signal and $\lfloor.\rceil$ represents the rounding function. This method achieves the same performance as in~\ref{nCoh_corr}, but at less processing complexity. However, when compared to coherent detection, non-coherent methods come at the cost of higher BER in AWGN environments.s. 

\subsection{Dataset Creation}
To create a synthetic dataset  of LoRa symbols, we use the open-source MATLAB emulator~\cite{9395074,Emulator} which was previously developed by our team. The target and interferer LoRa signals are generated as I/Q waveforms, with a $f_\mathrm{s} = 1$~MHz, a $\text{SF} = 7$, and a $B = 125$~kHz. A randomly generated LoRa message consisting of the symbol sequence vector ${M = \{m_\mathrm{1},m_\mathrm{2},...,m_\mathrm{n}\}}$ is utilized to create each waveform. For injecting interference to the original LoRa signal with a random phase shift, another LoRa signal, representing the interference, is added to the original signal which also has a random phase shift and delay. Finally, the resultant signal is resampled to Nyquist rate of $f_\mathrm{s} = 125$~kHz. Superimposing the target signal with interference while the two signals are sampled at a higher sampling rate allows for a more realistic signal due to the higher resolution.  
For the training and detection process, a sequence is picked up arbitrarily to represent the \emph{target} signal and is assigned a controlled power of $p_\mathrm{s}$. In addition to the interference, the received target signal is further impaired with a Gaussian noise process having a controlled power. We denote the controlled interference power as $p_\mathrm{I}$ and the noise power as $\sigma^2$. The INR is denoted as $\alpha$, and the SINR is donated as $\gamma$,
\begin{align}
\alpha = \frac{p_\mathrm{I}}{\sigma^2}, && \gamma = \frac{p_\mathrm{s}}{p_\mathrm{I}+\sigma^2} .
\end{align}
Since the performance is only related to the INR and SINR, the dataset consists of signals with a uniformly distributed SINR between $\{-15, 15\}$ dB and a uniformly distributed INR between $\{-30, 30\}$ dB. Additionally, we normalize the interference and noise power concerning the target LoRa signal power, i.e., $p_\mathrm{s}=1$. Accordingly, the stored emulated received waveform is comprised of three parts; (i) the target LoRa signal, (ii) the interference, (iii) the AWGN noise, as follows,
\begin{multline}
r(t) = \underbrace{ \vphantom{{\sqrt{\frac{\alpha}{\gamma 0+ \alpha\gamma}}}} x(t)\exp\left(j\phi_\mathrm{s})\right)}_{\text{Signal}} + 
\underbrace{\sqrt{\frac{\alpha}{\gamma + \alpha\gamma}} x_\mathrm{I}(t-\tau)\exp\left(j\phi_\mathrm{I}\right)}_{\text{Interference}} + \\
\underbrace{\sqrt{\frac{1}{\gamma + \alpha\gamma}} n(t)}_{\text{AWGN}},
\end{multline}
where $x(t)$ is the target LoRa baseband signal, $x_\mathrm{I}(t)$ is the interfering baseband LoRa signal with a random time shift of $\tau$, and $n(t) \sim \mathcal{CN}\left(0,\sigma^2\right)$ is the complex zero-mean AWGN. Both the target signal and the interfering signal have a uniformly distributed random phase shift between $0\degree~\text{and}~360\degree$ denoted by $\phi_\mathrm{s}$ and $\phi_\mathrm{I}$ respectively. Then, to normalize the signal, the overall signal is divided by the target signal power as follows,
\begin{equation}
    r_\mathrm{norm}(t)=\frac{r(t)}{\sqrt{p_\mathrm{s}}} .
\end{equation}

\begin{figure}[!t]
	{\centering
		\includegraphics[width=\linewidth]{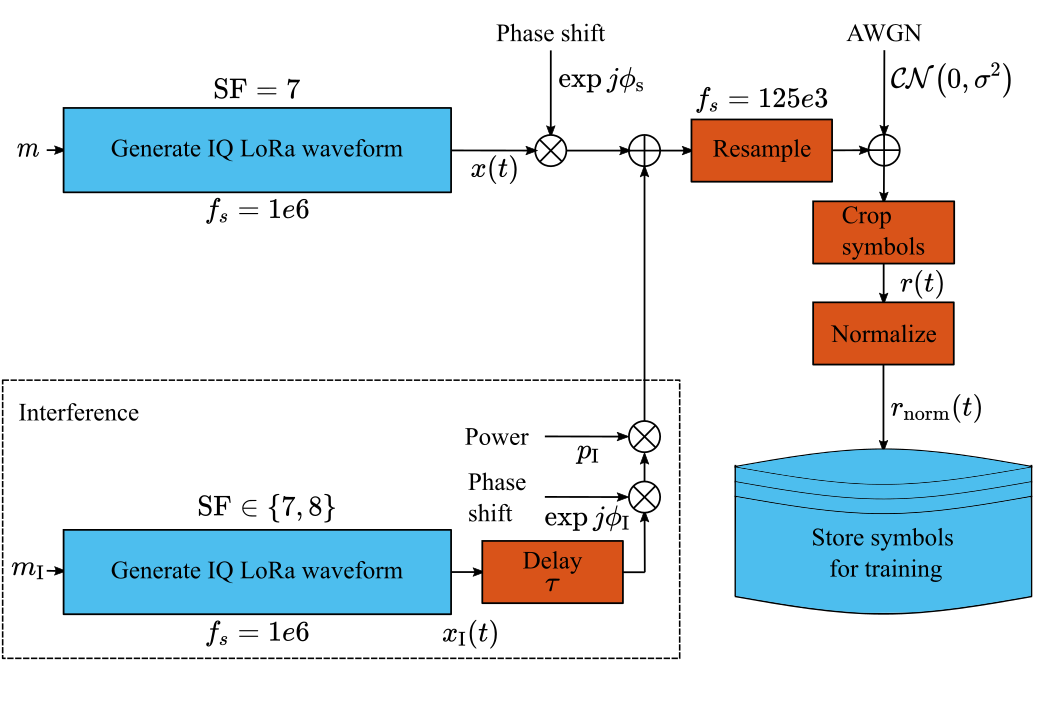}
        \caption{Dataset generation process for training each CNN. }
		\label{Fig_dataset}}
	\footnotesize
\end{figure}

The complex baseband signal is cropped into LoRa symbols, where each target symbol is synchronized in time, as such this implicitly assumes that a receiver is capable of performing preamble detection and synchronization with the target signal. Each symbol in the dataset is labeled according to the transmitted symbols. Since increasing the SF leads to more permissible values in a symbol (e.g. SF 7 means 128 different variations could be represented in a symbol, and SF 8 means 256, etc..), the training dataset size will increase due to the number of classes increasing according to the number of possible symbol values that a symbol can encode, which is equal to $2^\mathrm{SF}$. Consequently, the training time would be much longer, and the network complexity would also need to be increased. An illustration of the creation process of the dataset is depicted in Fig~\ref{Fig_dataset}. 

\begin{table}
 \caption{Notations and Symbols}
 \centering
 \begin{tabular}{l c c}
 	\hline\hline
 	Parameter & Symbol &Value \\
 	\hline
    Spreading factor & SF &  7  \\
    Bandwidth & $B$  & 125~kHz \\
    Sampling rate & $f_\mathrm{s}$ & 125~kHz \\
    Samples per symbol & $N$ & 128 \\
    Frequency offset & $\zeta$ & - \\
    Chirp slope & $\beta$ & - \\
    Symbol time & $T_\mathrm{s}$ & 1.024~ms \\
    Noise power spectral density & $N_{\mathrm{o}}$ & -\\
    Symbol value & $m$ & -\\
    Discrete frequency step & $\delta_{\text{f}}$ & 976.56~Hz \\
    Average noise power & $\sigma^2$ & - \\ 
    LoRa symbol & $s_{k}(t)$ & - \\
    Dechirping LoRa signal & $s^\star(t)$ & - \\                                                   
    Dechirped LoRa signal & $y(t)$ & - \\   
    FSK signal FFT & $Y(f)$ & -\\
    Non-coherent FSK symbol & $\hat{m}_{\text{n-coh}}$ & -\\    
    Coherent FSK symbol & $\hat{m}_{\text{coh}}$ & -\\    
    Interference to noise ratio (INR)  & $\alpha$ & -\\
    Signal to interference plus noise ratio (SINR) & $\gamma$ & -\\
    Target signal power & $p_\mathrm{s}$ & 1 \\
    Interference signal power & $p_\mathrm{I}$ & - \\
    Number of training frames &-& 400,000 \\
    Number of validation frames &-& 100,000 \\
    STFT window type & - & Hamming \\
    Number of DFT points & $N_\mathrm{DFT}$ & 128 \\
    Points per FFT window for the STFT & $W$ & 16 \\
    STFT window overlap & $L$ & 15 \\
    I/Q modality network input & $R$ & - \\
    Time-frequency modality network input & $X$ & - \\
    Spectrum modality network input & $F$ & - \\
 	\hline
 \end{tabular}
 \label{Table_LoRa}
\end{table}    

\section{Deep Learning Detection}~\label{Section_IV}
Following typical CNN architectures, the main elements that commonly make up a CNNs are as follows: (i) convolutional layers interlaced with (ii) max-pooling layers and (iii) batch normalization layers. The convolutional layer performs the convolution operation on the input with a kernel (filter) with a certain dimension. The window of the kernel slides across the input data with the a unity stride. The output is then multiplied by an activation function, where the Rectified Linear Unit (ReLU) is utilized. Note that the ReLU function is linear for positive inputs and gives zero for negative inputs. Typically, each convolutional layer is followed by a batch normalization layer, which normalizes the mean and variances of the convolutional layer output, which speeds up training. A max-pooling layer then follows, which down-samples the data by taking the maximum value in each max-pooling filter. The output is then passed into a fully connected layer that uses a Softmax activation function that outputs an $l$-length vector of scores summing to 1, where $l$ is the number of classes. Finally, a classification layer assigns the classes according to the probabilities.

\subsection{Data Modalities}\label{Sec_Modelities}
In this section, we explain the three different data modalities that could be used for detecting LoRa signals:

\subsubsection{I/Q Modality}
The first model is constructed with time-domain I/Q samples. The model takes in the complex input signal represented as the real part,
\begin{equation}
    R_\mathrm{I} = \mathfrak{Re}[r(1),r(2),...,r(N)] , 
\end{equation}
and the imaginary part,
\begin{equation}
    R_\mathrm{Q} = \mathfrak{Im}[r(1),r(2),...,r(N)] ,
\end{equation}
where for each time-domain symbol $r(.)$ there are $N$ temporal samples. The samples are then arranged into two 1D vectors to be used in the DL network as follows,
\begin{equation}
    R = \left[ \begin{array}{cc}
    R_\mathrm{I} \\
    R_\mathrm{Q} \end{array} \right] .
\end{equation}

\subsubsection{Time-frequency Domain Modality}
In the second modality, we convert the time domain samples into a spectrogram using the squared magnitude of the short-time Fourier transform (STFT). STFT works by taking segments, \emph{strides}, of the time domain signal and converting each one using the FFT. After that, the FFT vectors are combined in a 2D matrix representing the spectral change across time. In such representation, a linear chirp, for example, will appear as a straight line. The calculation of a spectrogram matrix is achieved by performing the STFT and then taking the squared magnitude as follows,
\begin{equation}
    X(\omega,p)=\left|\sum_{n=0}^{N_\mathrm{DFT}-1} r(n)g(n-pO)e^{-j\omega n}\right|^2 ,
\end{equation}
where $X(\omega,p) \in \mathbb{R}^{N_\mathrm{DFT} \times \frac{N - L}{W - L}}$. ${r(.)}$ denotes the captured sample of a cropped discrete-time LoRa baseband symbol. Every $p$th column of the spectrogram matrix has $\frac{N - L}{W - L}$ values. A Hamming windowing function is denoted by $g(.)$ with an overlap length between each Discrete Fourier Transform (DFT) of $O = W-L$. The parameters chosen for the STFT are outlined in Table~\ref{Table_LoRa}. 

\subsubsection{Frequency Domain Modality}
The third DL approach uses the frequency domain, where only the absolute of the FFT of a dechiped LoRa symbol is used. The input signal can be represented as follows,
\begin{equation}
    F(\omega) = \left|\sum_{n=0}^{N-1} r(n)e^{-j\omega n/N}\right| ,
\end{equation}
where $r(.)$ is the dechirped LoRa symbol and $N$ is the number of temporal samples. This modality exploits domain knowledge by pre-processing the input data to simplify the detection process. 

\subsection{Hybrid Architecture: HybNet}~\label{Section_HybNet}
The proposed HybNet architecture is the core contribution of this paper, which in essence is a classifier that switches between two different detection branches; (i) a DL network (in our case we select the FFT-CNN as referred to in Fig.~\ref{Fig_Networks}), due to the performance characteristics which are shown in Section~\ref{Section_V}, and (ii) a traditional matched detector (in our case we utilize non-coherent detection). This architecture is developed to marry the advantages of both detectors, i.e. the optimal performance of matched filters in AWGN and the improved performance of CNN in co-channel interference conditions. In order to decide which detector branch to choose, a supervisory selector system is utilized to decide whether to pass the signal to the first branch or the second. 

\begin{figure}[!t]
{\centering
\includegraphics[width=\linewidth]{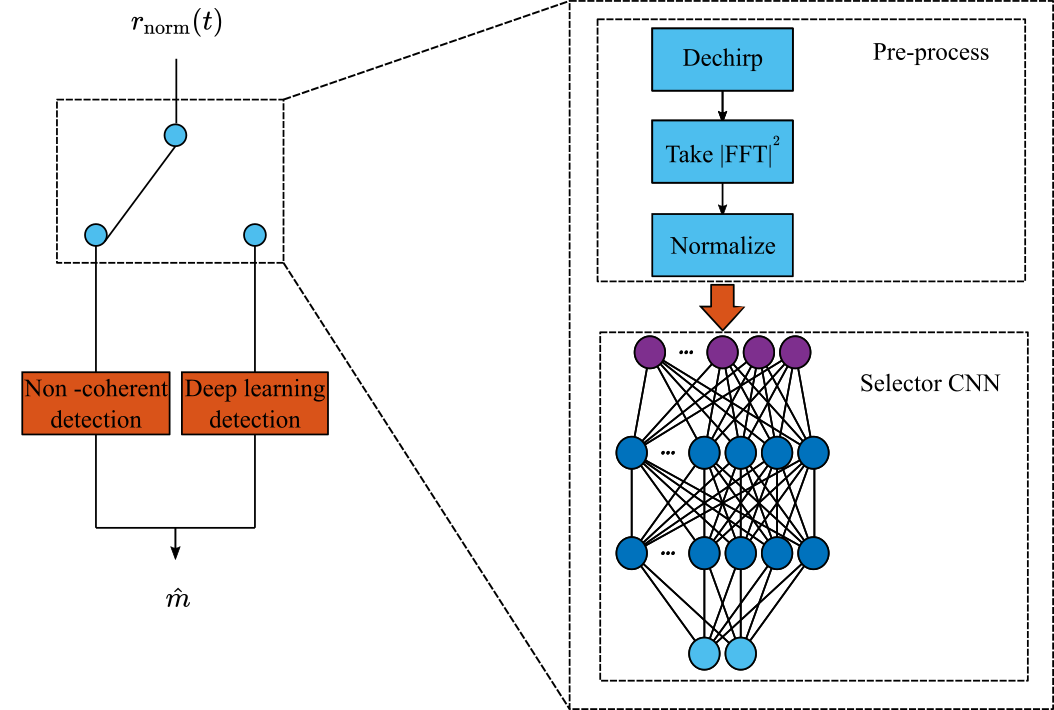}
\caption{Illustration of the proposed HybNet architecture switching between a deep-learning branch and a matched filter-branch based on the interference-to-noise ratio. }
\label{Fig_Models}}
\footnotesize
\end{figure}

Note that this selector is trained with another CNN network, which we dub as the \textit{Selector CNN}. A dataset consisting of 200,000 symbols with a uniformly distributed SINR ranging between $\{-15, 15\}$ dB and a uniformly distributed INR between $\{-30, 30\}$ dB is created to be trained on. Each LoRa symbol is labeled with either one of two possible classes; (i) \emph{Traditional} or (ii) \emph{Deep learning}. If the received symbol is identified to lack LoRa interference or the gain of the interfering LoRa signal is less than that of the target, the symbol is labeled as ``Traditional''. Consequently, the received symbol is passed to be purely detected by the non-coherent detector (outlined in subsection~\ref{nCoh_subsection}.) On the other hand, if the gain of the interfering transmissions is larger than the gain of the target, the symbol is labeled as ``Deep learning''. In that case, the received symbol is passed onto the DL detection branch. An illustration of the utilized switching architecture is shown in~Fig.~\ref{Fig_Models}. 

\begin{figure}[!t]
	{\centering
		\includegraphics[width=\linewidth]{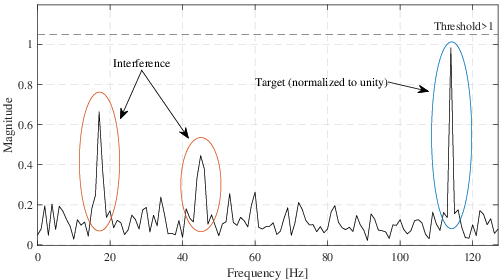}
		\caption{Illustration of how the energy detector switching works. }
		\label{ED_fig}}
	\footnotesize
\end{figure}

Instead of the Selector CNN, simple threshold switching can be used, whereby the simple switching method based on the energy level is shown in Fig.~\ref{ED_fig}. If the peak magnitude of the FFT exceeds the threshold, then the signal is passed to be detected by the DL detection branch. On the other hand, if the peak magnitude of the FFT falls below the threshold, non-coherent detection is used. 

\begin{figure}[!t]
	{\centering
		\includegraphics[width=\linewidth]{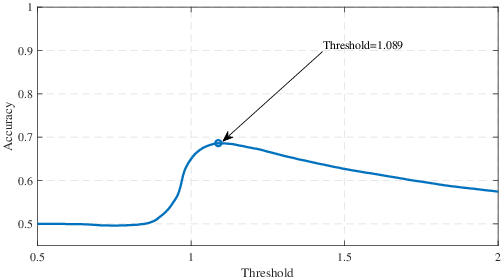}
		\caption{The plot of the overall accuracy for the energy detector against a varying threshold value. }
		\label{threshold_curve}}
	\footnotesize
\end{figure}

\begin{figure}[!t]
	{\centering
		\includegraphics[width=\linewidth]{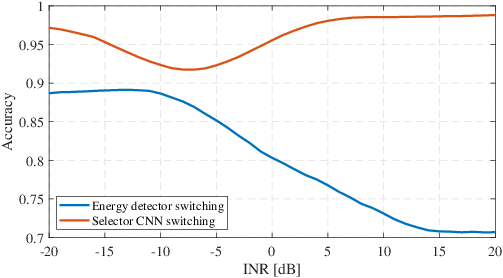}
		\caption{The plot of the overall accuracy against a varying INR value to compare the performance of a threshold energy detector and the Selector CNN. }
		\label{threshold_curve_2}}
	\footnotesize
\end{figure}

To evaluate the efficacy of a simple threshold energy detector instead of a DL network for the switching logic, different threshold values are accessed and plotted against the switching accuracy, which is shown in Fig.~\ref{threshold_curve}. The figure shows that the optimum accuracy of the energy detector occurs at a threshold of magnitude $1.089$. Note that the resulting FFT is normalized by $Y = \text{FFT}(y)/\text{length}(y)$. The performance of the energy detector is also compared to the Selector CNN switching network at varying INR levels in~Fig.~\ref{threshold_curve_2}. For this test, a dataset consisting of 20,000 symbols with a uniformly distributed SINR between $\{-15, 15\}$ dB per INR point is used. From the figure, the Selector CNN consistently classifies whether non-coherent or DL detection should be used to reduce the error rate from the detection process. On the other hand, threshold switching performs worse than the Selector CNN at all INR values. 

\subsection{CNN Setup Description}
\begin{figure}[!t]
{\centering
\includegraphics[width=\linewidth]{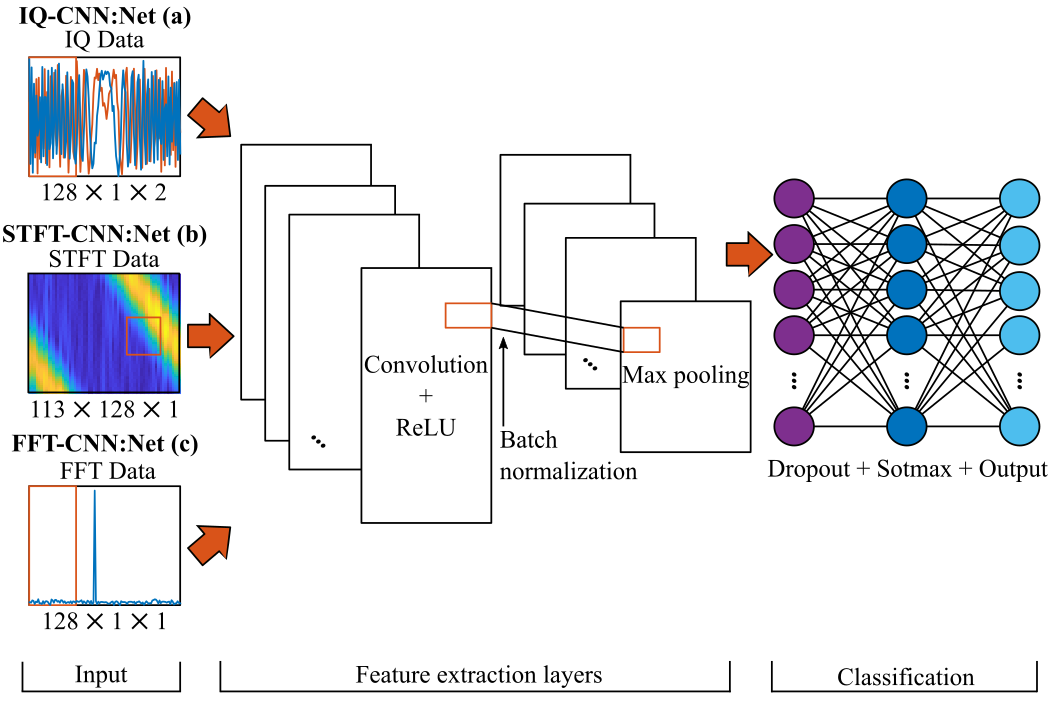}
\caption{Illustration of all three DL-based detector models used in this paper. }
\label{Fig_Networks}}
\footnotesize
\end{figure}

Three different CNN networks were designed to cater to the three data modalities explained in Section~\ref{Sec_Modelities}, and this is because both input dimensions and contents are different. For each of the three networks, we utilize a Bayesian optimizer to tune the network's hyperparameters. Bayesian optimization is a more efficient method for selecting hyperparameters compared to search methods such as brute-force, grid search, and random search~\cite{bayesian}. The hyperparameters that were chosen for optimization are; (i) the number of convolutional layers, (ii) convolutional filter size, (iii) initial learning rate, and (iv) the dropout rate. The resulting network architectures are summarized in Table~\ref{Table_CNN}. The summary of the CNN architectures are also illustrated in Fig~\ref{Fig_Networks}, showing the input format to each network. 

Considering the simple classification task, the network architectures are shallow. The Selector network, on the other hand, is manually optimized for the hard switching architecture since high accuracy is achieved with a straightforward network. A simple two-layer CNN is used with a convolutional layer width of 16 using the ReLU activation function. All the networks were trained with stochastic gradient descent with momentum (SGDM) over 60 epochs. The training parameters for all the networks discussed in this paper are outlined in Table~\ref{Table_trainingOpt}.

\begin{table*}\centering
 \caption{ CNN Network Summary }
 \centering
 \begin{tabular}{@{}lccccc@{}} 
 	\hline\hline
     {Layer} & \multicolumn{3}{c}{Shape} &
     \phantom{abc} & {Parameter} \\
     \cmidrule{2-4}
     & $\mathrm{IQ-CNN}$ & $\mathrm{STFT-CNN}$ & $\mathrm{FFT-CNN}$ \\ 
     & $\mathrm{Net(a)}$ & $\mathrm{Net(b)}$ & $\mathrm{Net(c)}$ \\
     \hline
        Input  & $128 \times 1 \times 2$ & $113 \times 128 \times 1$ & $128 \times 1 \times 1$ &&-\\
        Convolutional 1 & $57$, $19 \times 1$ & $57$, $7 \times 7$ & $57$, $11 \times 1$ && ReLU\\
        Batch Normalization 1 &- &- &- &&-\\
        Max Pooling 1 & $2 \times 1$ & $2 \times 1$ & $2 \times 1$ && - \\
        Convolutional 2 & $57$, $19 \times 1$ & $57$, $7 \times 7$ & $57$, $11 \times 1$ && ReLU\\
        Batch Normalization 2 &- &- &- &&-\\
        Max Pooling 2 & $2 \times 1$ & $2 \times 1$ & $2 \times 1$ && - \\
        Convolutional 3 & $57$, $19 \times 1$ & $57$, $7 \times 7$ & $57$, $11 \times 1$ && ReLU\\
        Batch Normalization 3 &- &- &- &&-\\
        Max Pooling 3 & $2 \times 1$ & $2 \times 1$ & $2 \times 1$ && - \\
        Convolutional 4 & $57$, $19 \times 1$ &  $57$, $7 \times 7$ & $57$, $11 \times 1$ && ReLU\\
        Batch Normalization 4 &- &- &- &&-\\
        Max pooling 4 & $2 \times 1$ & $2 \times 1$ & $2 \times 1$ && -\\
        Convolutional 5 & $57$, $19 \times 1$ &  $57$, $7 \times 7$ & $57$, $11 \times 1$ && ReLU\\
        Batch Normalization 5 &- &- &- &&-\\
        Max pooling 5 & $2 \times 1$ & $2 \times 1$ & $2 \times 1$ && -\\
        Dropout & $0.43$ & $0.37$ & $0.50$ && - \\
        Fully Connected & $128$ & $128$ & $128$ && Softmax \\
    \hline
 \end{tabular}
 \label{Table_CNN}
\end{table*}

\begin{table}
 \caption{Training Options}
 \centering
 \begin{tabular}{l c}
 	\hline\hline
 	Parameter & Value \\
 	\hline
 	Optimizer & {SGDM} \\
 	Momentum & {0.9} \\
 	Epochs & {60} \\
 	Learning rate drop schedule & {40} \\
 	Learning rate drop factor & {0.1} \\
    Mini-batch size & {256} \\
    L2 regularization & {0.0001} \\
 	\hline
 \end{tabular}
 \label{Table_trainingOpt}
\end{table}   

\section{Simulation Results and Discussion}~\label{Section_V}
This section presents the performance benchmark of the different architectures based on the BER indicator. A Monte-Carlo simulation of both \textit{same-SF} interference (SF7 on SF7) and \textit{inter-SF} interference (SF8 on SF7) is performed. The performance is investigated for a variable level of INR. A lower INR value indicates a noise-dominant scenario (noise-limited performance), while a higher value indicates an interference-dominant scenario (interference-limited performance). We have not presented BER plots for varying SINR because BER is known to enhance with the increasing SINR, a consistent trend in any detection algorithm. As such, we pick a transitional SINR where INR significantly impacts the BER performance, without loss of generality. Note that this work investigates the detection performance for signals with a normalized signal power $p_\mathrm{s}=1$ unit power. We assume that the receiver is able to estimate the received power of the target signal based on an uncorrupted preamble. This is because the focus of this work is on the retrieval of the corrupted payload which has an order of magnitude longer duration than the preamble.

\subsection{Detection Performance}
\begin{figure}[!t]
{\centering
\includegraphics[width=\linewidth]{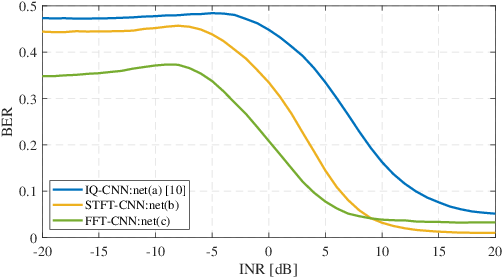}
\caption{Detection performance for a target LoRa symbol with SF7, an interference LoRa signal with SF7, and a fixed $\text{SINR} = -15 \text{dB}$. }
\label{Fig_ber_7_-15}}
\footnotesize
\end{figure}

\begin{figure}[!t]
{\centering
\includegraphics[width=\linewidth]{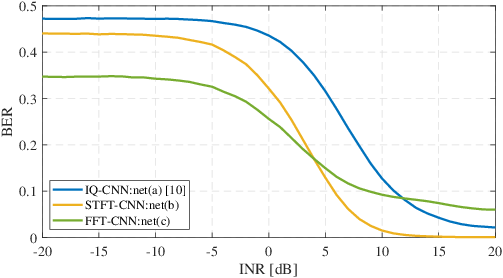}
\caption{Detection performance for a target LoRa symbol with SF7, and an interference LoRa signal with SF8, and a fixed $\text{SINR} = -15 \text{dB}$. }
\label{Fig_ber_8_-15}}
\footnotesize
\end{figure}

The BER performance of the DL-based techniques is depicted in Fig.~\ref{Fig_ber_8_-15} showing the performance in an \textit{inter-SF} interference scenario. Furthermore, we put the DL-based detectors to the test when we check the performance for a \textit{same-SF} LoRa interferer, with a SF7 in Fig.~\ref{Fig_ber_7_-15}. The figures both show that the FFT-CNN has the best performance under noise-limited scenarios. However, at high INR vales with same-SF interference the STFT-CNN very slightly outperforms the FFT-CNN. Although, at high INR values in an inter-SF scenario, both the IQ-CNN and STFT-CNN outperform the FFT-CNN. This is due to the FFT-CNN having no time information, unlike the IQ-CNN and STFT-CNN. 

\begin{figure}[!t]
{\centering
\includegraphics[width=\linewidth]{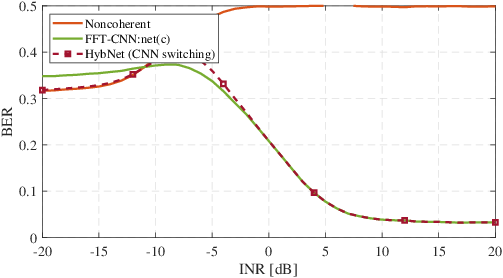}
\caption{Detection performance of the switching architecture for a target LoRa symbol with $\text{SF} = 7$, an interference LoRa signal with $\text{SF} = 7$, and a fixed $\text{SINR} = -15 \text{dB}$. }
\label{Fig_ber_7_-15_Hyb}}
\footnotesize
\end{figure}

\begin{figure}[!t]
{\centering
\includegraphics[width=\linewidth]{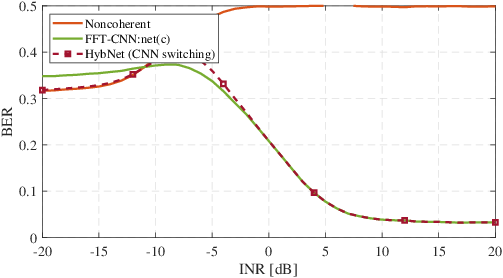}
\caption{Detection performance of the switching architecture for a target LoRa symbol with $\text{SF} = 7$, an interference LoRa signal with $\text{SF} = 8$, and a fixed $\text{SINR} = -15 \text{dB}$. }
\label{Fig_ber_8_-15_Hyb}}
\footnotesize
\end{figure}

\begin{figure}[!t]
{\centering
\includegraphics[width=\linewidth]{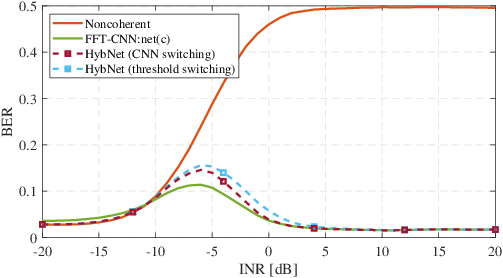}
\caption{Detection performance of both the CNN switching architecture and the energy detection switching architecture for a target LoRa symbol with $\text{SF} = 7$, an interference LoRa signal with $\text{SF} = 7$, and a fixed $\text{SINR} = -10 \text{dB}$. }
\label{Fig_ber_7_-10_Hyb}}
\footnotesize
\end{figure}

\begin{figure}[!t]
{\centering
\includegraphics[width=\linewidth]{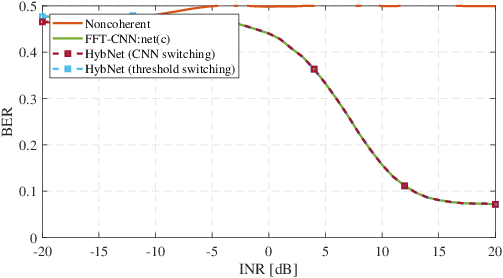}
\caption{Detection performance of both the CNN switching architecture and the energy detection switching architecture for a target LoRa symbol with $\text{SF} = 7$, an interference LoRa signal with $\text{SF} = 7$, and a fixed $\text{SINR} = -20 \text{dB}$. }
\label{Fig_ber_7_-20_Hyb}}
\footnotesize
\end{figure}

To present the performance improvement of the \emph{HybNet} framework compared to the DL-based detectors, fixed INR plots with a varying SINR are used. Traditional non-coherent methods (upper-bound) is plotted alongside the HybNet framework and FFT-CNN performance. The performance of HybNet is depicted in Fig.~\ref{Fig_ber_7_-15_Hyb} and in Fig.~\ref{Fig_ber_8_-15_Hyb} for SF7 and SF8 interference respectively, at a SINR at value of $\gamma=-15~\mathrm{dB}$. It can be clearly seen how the proposed HybNet architecture can effectively switch between the coherent detector and the FFT-CNN branches and thus follows the optimal performance in both noise-limited and interference-limited scenarios. The efficient switching indicates that the interference detector network can accurately select a detector depending on the interference level. Furthermore, the switching performance of the CNN compared to the energy detector can be observed in Fig.~\ref{Fig_ber_7_-10_Hyb}. From the plot, it can be seen that the energy detector switches between the non-coherent and DL pathways at a lower efficacy compared to the CNN. Also to note, the efficacy of the energy threshold switching network further degrades when the fixed SINR is decreased while the CNN switching network continues to be efficient. The performance of the HybNets at a fixed SINR of $\gamma=-20~\mathrm{dB}$ can be observed in Fig.~\ref{Fig_ber_7_-20_Hyb}. 

We further analyze the overall dataset accuracy, training time, and detection time of the three main architectures discussed in this paper, the IQ-CNN, STFT-CNN, and the FFT-CNN, in addition to the performance of HybNet and the HybNet architecture when the threshold switching (discussed in subsection~\ref{Section_HybNet}). We utilize MATLAB to create the testing dataset which consists of 200,000 symbols with a uniformly distributed SINR between $\{-15, 15\}$ dB and a uniformly distributed INR between $\{-30, 30\}$ dB. MATLAB is also used for prepossessing and for DL. The system used for the experimentation has a 16-logical core Intel Xeon CPU at 3.2 GHz and a Nvidia Quadro 4000 GPU. Table~\ref{Table_performance} shows the additional performance metrics of the networks, including the overall dataset accuracy, training time for each architecture, and detection time per symbol. 

For the purely DL detectors, the IQ-CNN requires the least amount of pre-processing compared to the other networks and has a low training time, however, it has the lowest accuracy. On the other hand, the STFT-CNN has higher accuracy, however, it also exhibits the longest training time and time per symbol. Notwithstanding, the FFT-CNN exhibits the lowest complexity, highest accuracy, and trains the fastest. The HybNet increases the dataset accuracy, however, the training time is slightly higher as it is the sum of FFT-CNN and the Selector CNN. To compromise between high performance and high complexity, the HybNet network can be used with the energy detector as the switching network instead of the Selector CNN, where the accuracy is higher than the non-coherent detector and the time per symbol is significantly lower than compared to the pure DL detectors. 

\begin{table}
	\caption{Performance analysis}
	\centering
	\begin{tabular}{l c c c}
		\hline\hline
		Network & Test accuracy & Train time & Time per symbol \\
		\hline
		 $\mathrm{Net(a)}$ & 78.66$\%$ & 12.2 min & 3.4$\times10^{-3}$ s \\
		 $\mathrm{Net(b)}$ & 84.84$\%$ & 140.75 min & 1$\times10^{-2}$ s \\
		 $\mathrm{Net(c)}$ & 86.74$\%$ & 12.2 min & 3$\times10^{-3}$ s \\
		 HybNet (CNN) & 87.68$\%$ & 14.5 min & 3.7$\times10^{-3}$ s \\
		 HybNet (threshold) & 81.8$\%$ & 12.2 min & 3.6$\times10^{-4}$ s \\ 
		 Non-coherent & 74.58$\%$ & - & 1.9$\times10^{-5}$ s \\
		\hline
	\end{tabular}
	\label{Table_performance}
\end{table}   

\subsection{Complexity Analysis}~\label{Section_complexity}
The theoretical time complexity for all the networks can be expressed as $O\left(L \sum_{l=1}^{M} K_{l-1} F_l W_l K_l\right)$~\cite{he2015convolutional}, where $l$ is the index of convolutional layers and $M$ is the number of layers. $L$ is the number of input symbols, and $K_{l-1}$ denotes the number of input channels. $F_{l}$ is the dimensions of the convolutional filter multiplied together. $W_{l}$ is the number of filters per convolutional layer. Finally, $K_{l}$ is the dimensions of the output multiplied together. The \textit{max-pooling}, \textit{dropout}, and \textit{fully connected layers} have insignificant time complexity compared to the convolutional layers, so their complexity is not included in the calculation. From the theoretical expression, the time complexity linearly increases with several input symbols for all three networks. The theoretical expression confirms that the time complexity is much higher for the STFT-CNN since the convolutional filters $F_{l}$ as well as the output dimensions $K_{l}$ are 2-dimensional. 

\section{Conclusion}~\label{Section_VI}
This paper investigated deep learning approaches that rely on convolutional neural networks for the detection of LoRa symbols in the presence of AWGN and colored interference. This paper presented a new framework HybNet that switches between traditional matched filter detection and deep learning detection. As such  the proposed architecture combines the merits of (i) the optimal detection in Gaussian noise based on the matched filter, with (ii) the improved performance of the deep learning detector under non-Gaussian interference. The paper tested different input data modalities for deep learning, namely; (i) I/Q-based, (ii) time-frequency-based, and (iii) spectrum-based. The spectrum-based deep detector showed the best detection performance in heavy interference conditions and the lowest time complexity compared to the I/Q-based and time-frequency-based networks. This improvement in the performance suggests that the proposed  hybrid architecture would outperform stand-alone conventional detection methods in a broader range of applications, especially in random access IoT networks where the interference is caused by overlapping co-channel transmissions.

\bibliographystyle{ieeetr}
\bibliography{bibliography}

\clearpage

\end{document}